\documentclass[prl,twocolumn,floatfix,nobibnotes,superscriptaddress]{revtex4}
\usepackage{amsmath}
\usepackage{graphicx}
\usepackage[tight]{subfigure}

\usepackage{color}

\begin{document}
\title{On the large magnetic anisotropy of Fe$_2$P}
\author{M. Costa}
\affiliation{Instituto de Fisica, Universidade Federal Fluminense, 24210-346 Niter\'oi, Rio de Janeiro, Brazil}
\affiliation{Department of Physics and Astronomy, Uppsala University, Box 516, 75121 Uppsala, Sweden}
\author{O. Gr\aa n\"as}
\affiliation{Department of Physics and Astronomy, Uppsala University, Box 516, 75121 Uppsala, Sweden}
\author{A. Bergman}
\affiliation{Department of Physics and Astronomy, Uppsala University, Box 516, 75121 Uppsala, Sweden}
\author{P. Venezuela}
\affiliation{Instituto de Fisica, Universidade Federal Fluminense, 24210-346 Niter\'oi, Rio de Janeiro, Brazil}
\author{P. Nordblad}
\affiliation{Department of Engineering Science, Uppsala University,Box 534, 75121, Uppsala, Sweden}
\author{M. Klintenberg}
\affiliation{Department of Physics and Astronomy, Uppsala University, Box 516, 75121 Uppsala, Sweden}
\author{O. Eriksson}
\affiliation{Department of Physics and Astronomy, Uppsala University, Box 516, 75121 Uppsala, Sweden}

\begin{abstract}
We present an investigation on the large magnetic anisotropy of Fe$_2$P, based on {\it Ab Initio} density-functional theory calculations, with a full potential linear muffin-tin orbital (FP-LMTO) basis.  We obtain an uniaxial magnetic anisotropy energy (MAE) of 664 $\mu$eV/f.u., which is in decent agreement with experimental observations. Based on a band structure analysis the microscopical origin of the large magnetic anisotropy is explained. We also show that by straining the crystal structure, the MAE can be enhanced further.
\end{abstract}
\pacs{later}

\maketitle

\section{Introduction}
Magnetic materials with large magnetic anisotropy (MAE) have been used in many applications. Until recently the dominating class of materials used for such applications, apart from hard ferrites, were rare-earth based magnets, e.g. Nd$_2$Fe$_{14}$B~\cite{ndfeb}, in which the large magnetic anisotropy energy is provided by the rare-earth atoms, and the large saturation moment (M$_s$) at finite temperature is due to the Fe atoms. It has however been pointed out that other permanent magnets should be investigated, from an application point of view, since a general access to rare-earth elements is far from guaranteed~\cite{physicstoday}. Among such materials the iron-phosphide Fe$_2$P stands out as a particularly interesting material, due its known large MAE and sufficient large value of saturation moment . In addition  Fe$_2$P is composed of cheap and widely available elements.

Transition metal pnictides and chalcogenides are not the only large MAE materials that have been investigated so far.
 For instance, a large saturation moment was suggested in a nano-laminate of a 3d metal (Fe) and a rare earth metal (Gd) in Ref.\onlinecite{heiko}. Also, a large magnetic anisotropy in FePt was verified experimentally as well as from first principles theory~\cite{FePt,FePt2,FePt3,staunton,WReFe}. Furthermore, the predicted large MAE of a tetragonally strained FeCo-alloy~\cite{burkert} was verified experimentally~\cite{gabi}. 

In this report we focus on the MAE of Fe$_2$P, since it is large and hitherto unexplained. In order to find a microscopic description of the large MAE of this compound we have performed first principles calculations, using a relativistic formulation of the Kohn-Sham equation. 
There are several experimental studies concerning the ordering temperature, MAE, saturation moment, hyperfine field and isomer-shift of Fe$_2$P~\cite{ref8,ref9,ref13,svane}. In addition, a theoretical analysis was made earlier by Wohlfarth~\cite{ref10a,ref10b}, and a subsequent theoretical work addressed the magnetism of Fe$_2$P using electronic structure calculations~\cite{ref11a,ref11b}. An excellent overview of the magnetic properties of Fe$_2$P and similar transition metal pnictides and chalcogenides can be found in Ref.\onlinecite{beckman}.

\begin{figure}[htb]
\begin{center}
\includegraphics[scale=0.35]{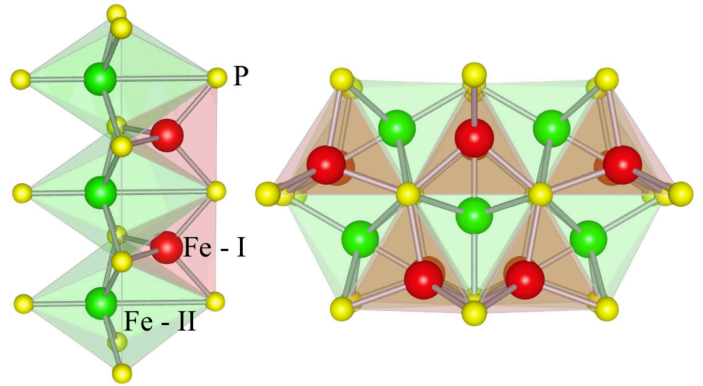}
\end{center}
\caption{\label{structure} (color online) Fe$_{2}$P crystal structure. Fe-I tetrahedral-sites (red), Fe-II pyramidal-sites (green) and P atoms (yellow).}
\end{figure}

In its crystal structure Fe$_2$P  (Hexagonal C$_{22}$ with space group P$\overline{6}$2m, No. 189)~\cite{carlsson,Klintenberg} has two Fe sites. The Fe-I site has a magnetic moment close to 1 $\mu_B$/atom, and the Fe-II site has a magnetic moment close to 2 $\mu_B$/atom (In Fig.~\ref{structure} the structure of Fe$_2$P is shown). The material has an observed MAE of 500 $\mu$eV/f.u. ~\cite{beckman}, with the crystallographic c-axis being the magnetization easy axis. The microscopic mechanism for this large MAE is however unknown. It is the motivation of the present study to find this mechanism, and to suggest Fe$_2$P based alloys with enhanced values of MAE. 

\makeatletter \renewcommand{\fnum@figure}
{Fig. \thefigure} \makeatother


\section {Computational Details}
We used a full potential linear muffin-tin (FP-LMTO) method~\cite{RSPt}, with a relativistic formulation, where the spin-orbit coupling is included inside the muffin-tins spheres, at the variational step. The muffin-tins are centered on atomic sites, and a combination of radial functions multiplied by spherical harmonics are used to expand the electronic density and potential inside each muffin-tin. In the interstitial region a combination of Hankel and Neumann functions are used as basis functions. We calculated the MAE by the magnetic force theorem~\cite{mft}. A scalar relativistic calculation, with common symmetry of both magnetization directions, is first performed to obtain a self-consistent potential. With this potential fixed one performs a fully-relativistic calculation for each magnetization orientations ($\hat n_{i}$). The E$_{MAE}$ (magnetic anisotropy energy) is then calculated from the expression
\begin{equation}
\label{mae}
E_{MAE} = \sum_{i,k}^{occ.} e_{i} (\hat n_{2},k) - \sum_{i,k}^{occ.} e_{i}(\hat n_{1},k).
\end{equation}
Here  $ e_{i} (\hat n,k)$ is the Kohn-Sham eigenvalues evaluated for each magnetization orientation. In Eq.1, $i$ labels the occupied states and k the {\bf k}-points in the Brillouin zone, whereas $\hat n_{1}$ = 0001 (c-axis) and $\hat n_{2}$ = 1000. 

The modified tetrahedron method (MTM) ~\cite{mtm} was used for integration in the Brillouin zone (BZ),  36002 {\bf k}-points were used (for a calculation with 50626 {\bf k}-points the MAE deviates only 0.1 \%), in irreducible part of BZ,  to guarantee the convergence.


\section {Results}

\begin{figure}
\centering
\includegraphics[scale=0.3,viewport=0 0 800 600]{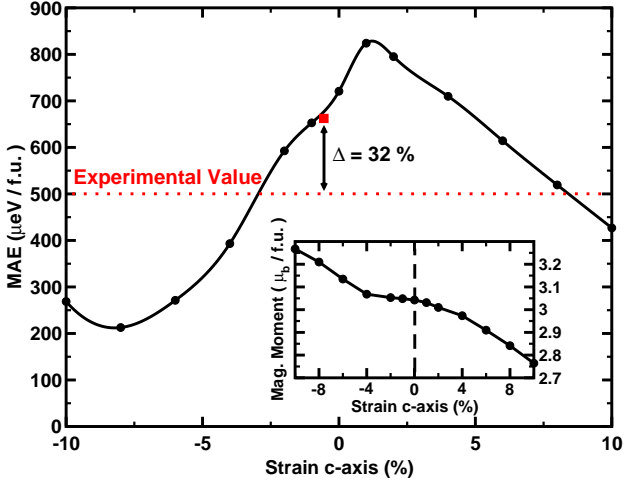}
\caption{\label{strain}(color online) Calculated MAE as function of the strain. Positive values of the strain correspond to an increased c/a ratio (at constant volume). Positive values of the MAE correspond to the c-axis being the easy axis. The inset shows the calculated magnetic moment as a function of strain. The square (red) is the MAE value for the estimated 4 K lattice parameters.}
\end{figure}

Our main result is shown in Fig.~\ref{strain}, where we display the calculated MAE as a function of the strain. Note that we have used the room temperature lattice constant as reference level, having room temperature applications in mind. It is clear from the figure that theory reproduces the observed easy axis (0001) and that the calculated MAE is of the same order of magnitude as the experimental one. The theory overestimates the value of the MAE by 32 \%, when a comparison is made between the experimental low temperature value (red square, taken at 4 K) and the lattice parameters corresponding to this temperature (which in Fig.2 corresponds to a - 0.55 \%). The strain state of the low temperature lattice constant is calculated from the work of Fujii {\it et. al.}~\cite{ref9} who reported measurements for the thermal expansion of the Fe$_{2}$P in the temperature range of 60 to 550 K. Using this thermal expansion data the 4 K lattice parameters were estimated. The agreement between calculated and measured MAE obtained here is typical, when compared to other calculations~\cite{FePt2,FePt3,trygg}, and primarily reflects the extremely delicate nature of the MAE in general. A very important result shown in Fig.~\ref{strain}, however, is that with an applied strain to the lattice it is possible to influence the MAE quite substantially. 
In these calculations, we applied strain, keeping a constant volume. Increasing the c value by 1\%  affects the MAE to be enhanced by $\sim$ 15 \%, which is the maximum value of the calculated MAE. For strain values higher than 1\%, Fig.~\ref{strain} shows that the MAE decreases almost linearly. Reducing the c-axis is not favorable for the MAE, here we note a monotonically decreasing trend, with a minimum observed at -8\%. Unfortunately most dopings on the P site, e.g. with Si or B, results in a reduced c/a ratio~\cite{ref9,fujii,carlsson}. 

The magnetic moment of Fe$_2$P changes much less with strain when compared to the MAE, as the inset of Fig.~\ref{strain} shows. At 1\% strain the calculated moment is 3.03 $\mu_B$/f.u., which is -0.4 \% of the zero strained system. The magnetic moment increases however slightly for negative values of strain. Overall our calculated moments agree well with the observed number of 2.94 $\mu_B$/f.u ~\cite{ref9}. 

\begin{figure}[htb]
\begin{center}
\includegraphics[scale=0.3,viewport=0 0 800 600]{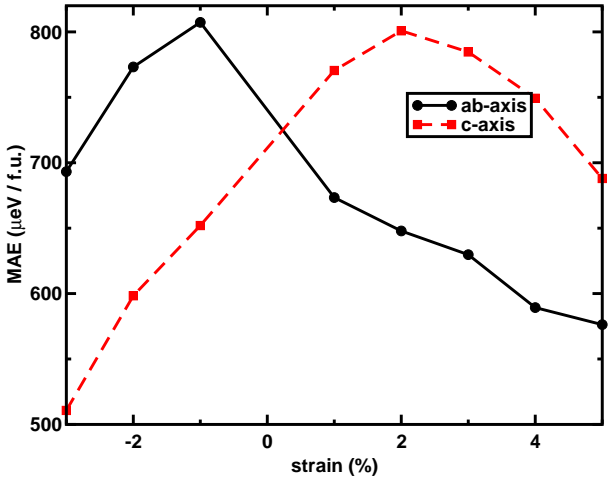}
\end{center}
\caption{\label{CxAB_strain} (color online) Calculated MAE as function of the strain of the c-axis with a and b fixed (red squares) and strain on the ab-axis with c fixed (black circles).}
\end{figure}

The Fig.~\ref{strain} shows data for a volume conserving strain (except the red square which corresponds to a volume which is minutely smaller than the volume used for the other data points). For comparison we also show in Fig.~\ref{CxAB_strain} the MAE as function of strain in two non-volume conserving regimes. 

This involves strain of the c-axis while a and b are fixed as well as strain of the ab-axis while c axis is fixed. For c-axis (ab-axis) strain a maximum MAE value of 800 $\mu$eV (807 $\mu$eV) is obtained for 2\% (-1\%) strain. Experimental doping on the P site with B results in a reduced c/a axis (negative strain) and a reduced volume~\cite{ref9,fujii,carlsson}, in this case a predicted MAE with doping is best evaluated by inspection of the non-volume conserving curve in Fig.~\ref{CxAB_strain}. Doping with Si keeps the volume essentially constant while reducing the c/a ratio, and for this doping element the predicted MAE is best evaluated from the volume conserving curve, which shows a decreased MAE. The MAE was actually calculated using the structural cell parameters for 10 \% Si doping~\cite{beckman}, and 593 $\mu$eV MAE was obtained, showing the expected reduction. For 10\% Si doping the experimental Curie temperature (T$_{c}$) is 370 K (an increase of 70 \% when compared with the undoped case), which is promising for stabilizing a material which at room temperature has a large MAE. Further investigations are necessary to consider the chemical effect impact on the MAE.

Both these predicted changes of the MAE with doping rely on the applicability of the rigid band approximation. They have not been evaluated experimentally, and a verification or refutal of this prediction would be interesting. 

\makeatletter \renewcommand{\fnum@figure}
{Fig. \thefigure} \makeatother

The strong variation of the MAE with strain in Fig.~\ref{strain} is interesting and requires further analysis. On a simple model level, the uniaxial MAE can be treated in second-order perturbation theory, and computed as the difference of the second order correction to the  energy ($E^{ss'}_{q}[\hat n]$)  between two magnetization directions as in Eq.~\ref{deltaE}, with a sum over the sites q, over occupied (s) and unoccupied (s') spin characters, respectively;~\cite{james}
\begin{equation}
\label{deltaE}
\Delta E_{SO} = \sum_{qss'} \Delta E^{ss'}_{q} = \sum_{qss'} \{E^{ss'}_{q} (\hat n_{2}) - E^{ss'}_{q} (\hat n_{1})\}, 
\end{equation}
where the energy correction is given by,

\begin{equation}
\label{E}
\begin{split}
E^{ss'}_{q} (\hat n) = -\sum_{{\bf k}ij}\sum_{q'} \sum_{\{m\}}  n_{{\bf k} is, qm,q'm'}n_{{\bf k} js', q'm'',qm'''} \times \\
\frac{\langle qms | \hat{H}_{SO} (\hat n) |  qm'''s' \rangle \langle q'm''s' | \hat{H}_{SO} (\hat n) |  q'm's \rangle}                                 {\epsilon_{{\bf k}j}- \epsilon_{{\bf k}i}}.
\end{split}
\end{equation}
In Eq.~\ref{E} there is a sum over {\bf k} points in the Brillouin zone, i and j label the occupied and unoccupied states, s and s' run over the spin character of the states and m, m$'$, m$''$, m$'''$ run over the magnetic quantum numbers. The basis functions $ | q l m s\rangle$ are characterized by the site q, azimuthal (l), magnetic (m) and spin (s) quantum numbers and $\epsilon_{{\bf k}i}$ ($\epsilon_{{\bf k}j}$) are the electronic eigenvalues for the occupied (unoccupied) states. The hybridization is considered in the band character $n_{{\bf k} is, qm,q'm'}$, which allows mixing of basis functions on different sites. 
\makeatletter \renewcommand{\fnum@figure}
{Fig. \thefigure} \makeatother
\begin{figure}[htb]
\begin{center}
\includegraphics[scale=0.3,viewport=0 0 800 600]{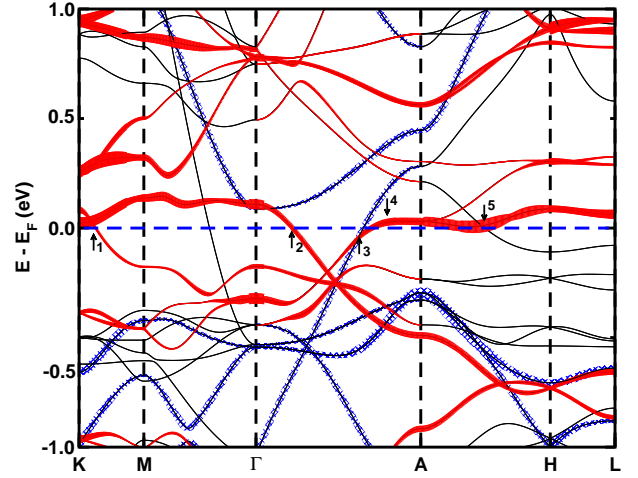}
\end{center}
\caption{\label{fatBands} (color online) Band structure of Fe$_{2}$P calculated without spin-orbit coupling (the band thickness represents the weight of Fe-II, l=2 and m=-2 state). Red (solid) is the spin up character and blue (striped) is the spin down. The Fermi level (E$_{F}$) is placed at E=0.}
\end{figure}

\makeatletter \renewcommand{\fnum@figure}
{Fig. \thefigure} \makeatother
\begin{figure}[htb]
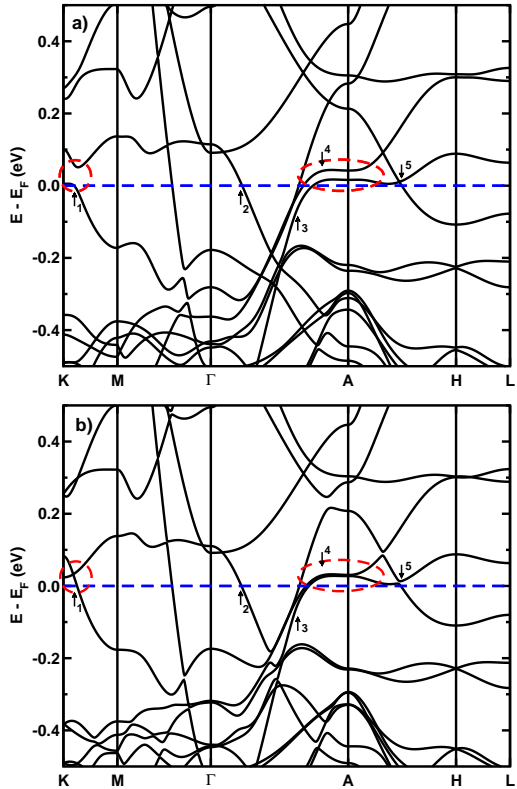

\begin{center}
\includegraphics[scale=0.25,viewport=0 0 800 600]{fig5_a.eps}
\includegraphics[scale=0.25,viewport=0 0 800 600]{fig5_b.eps}

\end{center}
\caption{\label{bands_soc} (color online) Zero strain band structure of Fe$_{2}$P calculated with spin-orbit coupling, magnetization along the a) 001 and b) 100 crystallographic axis. The Fermi level (E$_{F}$) is placed at E=0.}
\end{figure}

Since the electronic eigenvalues ($\epsilon_{{\bf k}i}$) appear in Eq.~\ref{E}, it is relevant
to inspect the band structure along the high symmetry lines of the hexagonal lattice. Hence, we show the calculated energy bands in Fig.~\ref{fatBands}, where the thickness of the bands represents the weight of the Fe-II, l=2 and m=-2 state. The bands that contribute the most to the MAE are highlighted by the arrows 1 to 5. At $\frac{1}{4}$(K-M), arrow 1, the occupied and unoccupied bands have mainly character from l=2 and m=$\pm$2 quantum numbers, and these bands interact through the $l_{z}s_{z}$ term of the spin-orbit coupling (SOC) Hamiltonian, resulting in a large negative contribution. A similar mechanism is observed at $\frac{3}{4}$($\Gamma$-A), arrow 4. This is illustrated further using a calculation with spin-orbit coupling included, in Fig.~\ref{bands_soc} a), in a region zoomed in around the Fermi level, see the highlighted areas. One can see the splitting of the bands when the spin quantization axis is along the 0001 crystallographic direction. The splitting of these bands is not observed for the 100 axis, as observed in Fig.~\ref{bands_soc} b). At arrows 2, 3 and 5 the occupied and unoccupied bands have different m quantum numbers, m=$\pm$2 and m=0 (m=$\pm$1) for the occupied (unoccupied) bands, the SOC interaction via the terms  : $l_{+}s_{-}$+$l_{-}s_{+}$, give large positive contributions. As a general rule if the occupied and unoccupied bands have the same (different) m quantum number, the magnetization is favored to lie along the 001 (100) axis~\cite{wu}.

\makeatletter \renewcommand{\fnum@figure}
{Fig. \thefigure} \makeatother
\begin{figure}[htb]
\begin{center}
\includegraphics[scale=0.3,viewport=0 0 800 600]{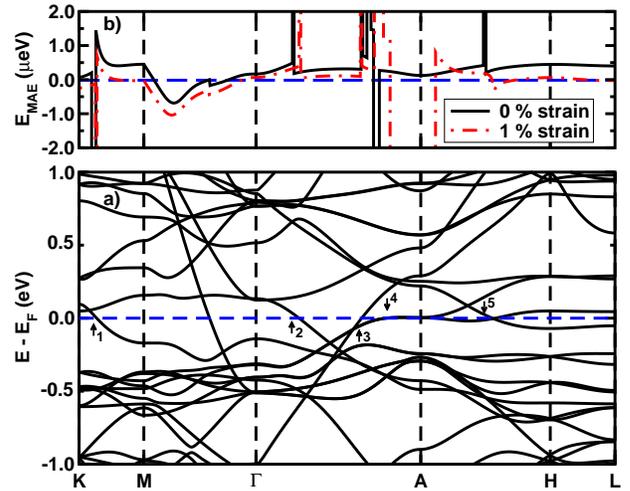}
\end{center}
\caption{\label{bandsMAE} (color online) a) Band structure of Fe$_{2}$P calculated with 1 \% of strain on the c-axis and b) MAE along the high symmetry directions. The Fermi Level (E$_{F}$) is placed at E=0.}
\end{figure}

To understand the enhancement of the MAE under uniaxial strain, the bandstructure  was calculated for the 1 \% volume conserving strain (Fig~\ref{bandsMAE} a). The overall bandstructure is not changed, due to the strain, only a rigid shift is observed. Without SOC the flat bands around the A point lies on the Fermi level. As pointed out before these bands are split due to SOC for a magnetization along the 001 axis. One of the bands become fully occupied and the other fully unoccupied, given that these bands are very close in energy a large contribution to the MAE is expected. Indeed  this is confirmed in Fig~\ref{bandsMAE} b) (upper panel) where the MAE for each {\bf k}-point (E$_{MAE}$[{\bf k}]) along the high symmetry lines of the hexagonal cell is showed, for the zero and 1\% (volume conserving) strained system. For the zero strain one can see a sharp negative peak for the MAE (negative values favors the 001 axis) around arrows 1 and 4. For 1\% strain the peak at arrow 4 gets broadened explaining the increasing of the MAE.

\section {Conclusion}
In conclusion we have studied the uniaxial MAE of Fe$_{2}$P by {\it Ab Initio} calculations. Our theory reproduces the observed 0001-easy axis with a MAE of 664 $\mu$eV/f.u. This should be compared to an experimental value of $\sim$ 500 $\mu$eV/f.u.(2.32 MJ/m$^{3}$) This is an acceptable agreement between theory and experiment, when having in mind the extremely delicate nature of the MAE and the typical low energy differences associated with it. The size of the MAE of Fe$_{2}$P should be compared to other hard magnetic materials like FePt 1.2 meV/f.u. (6.6 MJ/m$^{3}$)~\cite{FePt} and Nd$_2$Fe$_{12}$B 6.7 meV/f.u. (4.9 MJ/m$^{3}$)~\cite{Nd2Fe14B}.
As to the magnetic moments our calculations give value of 3.04 $\mu_B$/f.u., which agrees well with the experimental moment of 2.94 $\mu_B$/f.u.

We have analyzed the origin of the MAE by a detailed band and {\bf k}-point resolved property, and show that the positions of the different energy bands around the Fermi level critically determine that MAE. Since these bands can be moved up or down in energy with an applied strain, it seems that this is an important avenue with which to influence the MAE of Fe$_{2}$P and Fe$_{2}$P-based alloys. Consequently we find from our MAE calculations that it is possible to influence the MAE quite strongly with an applied strain, both in a volume conserving and a non-volume conserving mode. This opens up possibilities to use Fe$_{2}$P and alloys of this material, as a platform for searching for new permanent magnetic materials that don't contain rare-earth elements.

Support by the Swedish Research Council (VR) is acknowledged, as is
the the European Research Council (project 247062 - ASD),
the KAW foundation, eSSENCE, STEM, STANDUPP, EU project REFREEMAG and the Swedish National Infrastructure for Computing (SNIC). Valuable discussions with Laszlo Szunyogh are acknowledged.

\bibliographystyle{unsrtnat}
\bibliography{references}

\begin{thebibliography}{29}
\providecommand{\natexlab}[1]{#1}
\providecommand{\url}[1]{\texttt{#1}}
\expandafter\ifx\csname urlstyle\endcsname\relax
  \providecommand{\doi}[1]{doi: #1}\else
  \providecommand{\doi}{doi: \begingroup \urlstyle{rm}\Url}\fi

\bibitem[Coey(1996)]{ndfeb}
J.~M.~D. Coey, editor.
\newblock \emph{Rare-Earth Iron Permanent Magnets}.
\newblock Oxford University Press, 1996.

\bibitem[Kramer(2010)]{physicstoday}
David Kramer.
\newblock {Concern grows over China's dominance of rare-earth metals}.
\newblock \emph{Phys. Today}, 63\penalty0 (5):\penalty0 22, 2010.

\bibitem[Sanyal et~al.(2010)Sanyal, Antoniak, Burkert, Krumme, Warland,
  Stromberg, Praetorius, Fauth, Wende, and Eriksson]{heiko}
Biplab Sanyal, Carolin Antoniak, Till Burkert, Bernhard Krumme, Anne Warland,
  Frank Stromberg, Christian Praetorius, Kai Fauth, Heiko Wende, and Olle
  Eriksson.
\newblock {Forcing Ferromagnetic Coupling Between Rare-Earth-Metal and 3d
  Ferromagnetic Films}.
\newblock \emph{Phys. Rev. Lett.}, 104\penalty0 (15), April 2010.

\bibitem[Sun et~al.(2000)Sun, Murray, Weller, Folks, and Moser]{FePt}
Shouheng Sun, C.~B. Murray, Dieter Weller, Liesl Folks, and Andreas Moser.
\newblock Monodisperse fept nanoparticles and ferromagnetic fept nanocrystal
  superlattices.
\newblock \emph{Science}, 287\penalty0 (5460):\penalty0 1989--1992, 2000.
\newblock \doi{10.1126/science.287.5460.1989}.
\newblock URL \url{http://www.sciencemag.org/content/287/5460/1989.abstract}.

\bibitem[Daalderop et~al.(1991)Daalderop, Kelly, and Schuurmans]{FePt2}
Gho Daalderop, PJ~Kelly, and MFH Schuurmans.
\newblock {Magnetocrystalline Anisotropy and Orbital Moments in
  Transition-Metal Compounds}.
\newblock \emph{Phys. Rev. B}, 44\penalty0 (21):\penalty0 12054--12057, 1991.

\bibitem[Burkert et~al.(2005)Burkert, Eriksson, Simak, Ruban, Sanyal,
  Nordstr{\"o}m, and Wills]{FePt3}
Till Burkert, Olle Eriksson, Sergei Simak, Andrei Ruban, Biplab Sanyal, Lars
  Nordstr{\"o}m, and John Wills.
\newblock {Magnetic anisotropy of L10 FePt and Fe(1-x)Mn(x)Pt}.
\newblock \emph{Phys. Rev. B}, 71\penalty0 (13), April 2005.

\bibitem[Staunton et~al.(2004)Staunton, Ostanin, Razee, Gyorffy, Szunyogh,
  Ginatempo, and Bruno]{staunton}
J~B Staunton, S~Ostanin, S~S~A Razee, B~Gyorffy, L~Szunyogh, B~Ginatempo, and
  Ezio Bruno.
\newblock {Long-range chemical order effects upon the magnetic anisotropy of
  FePt alloys from an ab initioelectronic structure theory}.
\newblock \emph{J. Phys.: Condens. Matter}, 16\penalty0 (48):\penalty0
  S5623--S5631, November 2004.

\bibitem[Bhandary et~al.(2011)Bhandary, Gr{\aa}n{\"a}s, Szunyogh, Sanyal,
  Nordstr{\"o}m, and Eriksson]{WReFe}
Sumanta Bhandary, Oscar Gr{\aa}n{\"a}s, Laszlo Szunyogh, Biplab Sanyal, Lars
  Nordstr{\"o}m, and Olle Eriksson.
\newblock {Route towards finding large magnetic anisotropy in nanocomposites:
  Application to a W{\_}{(1-x)}Re{\_}{(x)}/Fe multilayer}.
\newblock \emph{Phys. Rev. B}, 84\penalty0 (9), September 2011.

\bibitem[Burkert et~al.(2004)Burkert, Nordstr{\"o}m, Eriksson, and
  Heinonen]{burkert}
Till Burkert, Lars Nordstr{\"o}m, Olle Eriksson, and Olle Heinonen.
\newblock {Giant Magnetic Anisotropy in Tetragonal FeCo Alloys}.
\newblock \emph{Phys. Rev. Lett.}, 93\penalty0 (2), July 2004.

\bibitem[Andersson et~al.(2006)Andersson, Burkert, Warnicke, Bj{\"o}rck,
  Sanyal, Chacon, Zlotea, Nordstr{\"o}m, Nordblad, and Eriksson]{gabi}
Gabriella Andersson, Till Burkert, Peter Warnicke, Matts Bj{\"o}rck, Biplab
  Sanyal, Cyril Chacon, Claudia Zlotea, Lars Nordstr{\"o}m, Per Nordblad, and
  Olle Eriksson.
\newblock {Perpendicular Magnetocrystalline Anisotropy in Tetragonally
  Distorted Fe-Co Alloys}.
\newblock \emph{Phys. Rev. Lett.}, 96\penalty0 (3), January 2006.

\bibitem[Wappling et~al.(1975)Wappling, H{\"a}ggstr{\"o}m, Ericsson,
  Devanarayanan, Karlsson, Carlsson, and Rundqvist]{ref8}
R~Wappling, L~H{\"a}ggstr{\"o}m, T~Ericsson, S~Devanarayanan, E~Karlsson,
  B~Carlsson, and S.~Rundqvist.
\newblock {First-Order Magnetic Transition, Magnetic-Structure, and Vacancy
  Distribution in Fe2p}.
\newblock \emph{J Solid State Chem}, 13\penalty0 (3):\penalty0 258--271, 1975.

\bibitem[Fujji et~al.(1977)Fujji, Hokabe, Kamigaichi, and Okamoto]{ref9}
H~Fujji, T~Hokabe, T~Kamigaichi, and T~Okamoto.
\newblock {Magnetic-Properties of Fe2p Single-Crystal}.
\newblock \emph{J. Phys. Soc. Jpn.}, 43\penalty0 (1):\penalty0 41--46, 1977.

\bibitem[Chandra et~al.(1980)Chandra, Bjarman, and ERICSSON]{ref13}
R~Chandra, S~Bjarman, and T~ERICSSON.
\newblock {ScienceDirect.com - Journal of Solid State Chemistry - A
  M{\"o}ssbauer and X-ray study of Fe2P(1-x)B(x) compounds (x < 0.15)}.
\newblock \emph{Journal of Solid State {\ldots}}, 1980.

\bibitem[Eriksson and Svane(1989)]{svane}
O~Eriksson and A~Svane.
\newblock {Isomer shifts and hyperfine fields in iron compounds}.
\newblock \emph{J. Phys.: Condens. Matter}, 1:\penalty0 1589, 1989.

\bibitem[Wohlfarth(1979)]{ref10a}
E~P Wohlfarth.
\newblock {First and second order transitions in some metallic ferromagnets}.
\newblock \emph{J. Appl. Phys.}, 50\penalty0 (B11):\penalty0 7542, 1979.

\bibitem[Moriya and Usami(1977)]{ref10b}
T.~Moriya and K.~Usami.
\newblock {Coexistence of ferro-and antiferromagnetism and phase transitions in
  itinerant electron systems}.
\newblock \emph{Solid State Communications}, 23\penalty0 (12):\penalty0
  935--938, 1977.

\bibitem[Ishida et~al.(1987)Ishida, Asano, and Ishida]{ref11a}
S~Ishida, S~Asano, and J~Ishida.
\newblock {Electronic-Structures and Magnetic-Properties of Mn2p, Fe2p, Ni2p}.
\newblock \emph{J Phys F Met Phys}, 17\penalty0 (2):\penalty0 475--482, 1987.

\bibitem[Eriksson et~al.(1988)Eriksson, Sj{\"o}str{\"o}m, Johansson,
  H{\"a}ggstr{\"o}m, and Skriver]{ref11b}
O~Eriksson, J~Sj{\"o}str{\"o}m, B~Johansson, L~H{\"a}ggstr{\"o}m, and
  HL~Skriver.
\newblock {Itinerant ferromagnetism in Fe2P}.
\newblock \emph{J Magn Magn Mater}, 74\penalty0 (3):\penalty0 347--358, 1988.

\bibitem[Buschow(1991)]{beckman}
K.H.J. Buschow, editor.
\newblock volume~6 of \emph{Handbook of Magnetic Materials}.
\newblock Elsevier, 1991.
\newblock \doi{10.1016/S1567-2719(05)80052-6}.
\newblock URL
  \url{http://www.sciencedirect.com/science/article/pii/S1567271905800526}.

\bibitem[Carlsson et~al.(1973)Carlsson, G{\"o}lin, and Rundqvist]{carlsson}
Bertil Carlsson, Margareta G{\"o}lin, and Stig Rundqvist.
\newblock {Determination of the homogeneity range and refinement of the crystal
  structure of Fe2P}.
\newblock \emph{J Solid State Chem}, 8\penalty0 (1):\penalty0 57--67, September
  1973.

\bibitem[Ortiz et~al.(2009)Ortiz, Eriksson, and Klintenberg]{Klintenberg}
C~Ortiz, O~Eriksson, and M~Klintenberg.
\newblock {Data mining and accelerated electronic structure theory as a tool in
  the search for new functional materials}.
\newblock \emph{Computational Materials Science}, 44\penalty0 (4):\penalty0
  1042--1049, February 2009.

\bibitem[Wills et~al.(2010)Wills, Alouani, Andersson, Delin, Eriksson, and
  Grechnev]{RSPt}
J~M Wills, M~Alouani, P~Andersson, A~Delin, O~Eriksson, and A~Grechnev.
\newblock \emph{Full-Potential Electronic Structure Method, Energy and Force
  Calculations with Density Functional and Dynamical Mean Field Theory}.
\newblock Springer Series in Solid-State Sciences. Springer, Berlin ;
  Heidelberg ; New York, 2010.

\bibitem[Jansen(1999)]{mft}
HJF Jansen.
\newblock {Magnetic anisotropy in density-functional theory}.
\newblock \emph{Phys. Rev. B}, 59\penalty0 (7):\penalty0 4699--4707, 1999.

\bibitem[Bl\"ochl et~al.(1994)Bl\"ochl, Jepsen, and Andersen]{mtm}
PE~Bl\"ochl, O~Jepsen, and OK~Andersen.
\newblock {Improved Tetrahedron Method for Brillouin-Zone Integrations}.
\newblock \emph{Phys. Rev. B}, 49\penalty0 (23):\penalty0 16223--16233, 1994.

\bibitem[Trygg et~al.(1995)Trygg, Johansson, Eriksson, and Wills]{trygg}
J~Trygg, B~Johansson, O~Eriksson, and JM~Wills.
\newblock {Total-Energy Calculation of the Magnetocrystalline Anisotropy Energy
  in the Ferromagnetic 3d Metals}.
\newblock \emph{Phys. Rev. Lett.}, 75\penalty0 (15):\penalty0 2871--2874, 1995.

\bibitem[Fujii et~al.(1979)Fujii, Komura, Takeda, Okamoto, Ito, and
  Akimitsu]{fujii}
Hironobu Fujii, Shigehiro Komura, Takayoshi Takeda, Tetsuhiko Okamoto, Yuji
  Ito, and Jun Akimitsu.
\newblock Polarized neutron diffraction study of fe$_{2}$p single crystal.
\newblock \emph{Journal of the Physical Society of Japan}, 46\penalty0
  (5):\penalty0 1616--1621, 1979.
\newblock \doi{10.1143/JPSJ.46.1616}.
\newblock URL \url{http://jpsj.ipap.jp/link?JPSJ/46/1616/}.

\bibitem[Andersson et~al.(2007)Andersson, Sanyal, Eriksson, Nordstr{\"o}m,
  Karis, Arvanitis, Konishi, Holub-Krappe, and Dunn]{james}
C~Andersson, B~Sanyal, O~Eriksson, L~Nordstr{\"o}m, O~Karis, D~Arvanitis,
  T~Konishi, E~Holub-Krappe, and J~Dunn.
\newblock {Influence of Ligand States on the Relationship between Orbital
  Moment and Magnetocrystalline Anisotropy}.
\newblock \emph{Phys. Rev. Lett.}, 99\penalty0 (17), October 2007.

\bibitem[Wang et~al.(1993)Wang, Wu, and Freeman]{wu}
D~Wang, R~Wu, and AJ~Freeman.
\newblock {First-principles theory of surface magnetocrystalline anisotropy and
  the diatomic-pair model}.
\newblock \emph{Phys. Rev. B}, 47\penalty0 (22):\penalty0 14932, 1993.

\bibitem[Abache and Oesterreicher(1986)]{Nd2Fe14B}
C~Abache and J~Oesterreicher.
\newblock {Magnetic anisotropies and spin reorientations of R2Fe14B-type
  compounds}.
\newblock \emph{J. Appl. Phys.}, 60\penalty0 (10):\penalty0 3671, 1986.

\end{thebibliography}
 
\end{document}